\documentstyle[12pt]{article}

\def\be{\begin{equation}}
\def\ee{\end{equation}}
\def\ben{\begin{displaymath}}
\def\een{\end{displaymath}}
\def\ba{\begin{array}{c}}
\def\ea{\end{array}}

\begin{document}

\titlepage
\vspace*{2cm}

 \begin{center}{\Large \bf
Unusual scalar products in Hilbert space of Quantum Mechanics:
 non-Hermitian square-well model with two coupled channels

 \vspace{.6cm}

   }\end{center}

\vspace{10mm}

 \begin{center}
Miloslav Znojil

 \vspace{3mm}

\'{U}stav jadern\'e fyziky AV \v{C}R, 250 68 \v{R}e\v{z}, Czech
Republic\footnote{e-mail: znojil@ujf.cas.cz}

\end{center}

\vspace{5mm}

%\today, albev.tex

\section*{Abstract}

A pseudo-Hermitian square-well model ``with spin" is proposed,
solved and discussed. The domain of parameters is determined where
all the bound-state energies remain real and where the necessary
transition from the original elementary non-physical indefinite
pseudo-metric to another, more involved but correct
positive-definite physical metric is possible.

\vspace{5mm}

PACS

03.65.Ge
% Solutions of wave equations: bound states

03.65.Ca
% Formalism

03.65.Ta
% Foundations of quantum mechanics; measurement theory

02.60.Lj
% Ordinary and partial differential equations; boundary
% value problems

\newpage

\section{Introduction}

Qualitative phenomenological description of quantum phenomena
often relies on a drastically simplified model. Most typically,
one derives an ``effective" reduced Hamiltonian $H_{\it e\!ff}$
from a given microscopic $H_{orig}$ by some approximation
procedure. In one of applications of such an approach in nuclear
physics, people map the fermions governed by a complicated
Hamiltonian $H_{orig}$) on a {\it bona fide} equivalent bosonic
system controlled by $H_{\it e\!ff}$ \cite{Geyer}.

Even though the latter recipe may lead to non-Hermitian $H_{\it
e\!ff}$ in general, the authors of the review~\cite{Geyer}
emphasized that the work with $H_{\it e\!ff} \neq H_{\it
e\!ff}^\dagger$ need not contradict the principles of Quantum
Mechanics. A core of the message lies in the observation that one
is free to introduce an alternative metric $\Theta \neq I$ in
Hilbert space. This operator defines the new scalar product and
the new norm in our Hilbert space in such a way that
 \be
 H_{\it e\!ff}^\dagger=\Theta\,
 H_{\it e\!ff}\,\Theta^{-1}\,,
 \ \ \ \ \ \ \Theta=\Theta^\dagger > 0\, .
 \label{quasihermitian}
 \ee
In this language, our Hamiltonian may be called ``quasi-Hermitian"
(the word meaning just Hermitian and, hence, physical with respect
to the new metric). It becomes allowed to represent an observable
(energy). Let us only note that it is in fact just conventional to
call $H$ ``the operator of energy" since it may also be
interpreted as ``an occupation number", etc. After all, the latter
re-classification is not too unusual, say, within the so called
superymmetric quantum mechanics \cite{Khare}.

In the late nineties, a renewed interest in the quasi-Hermitian
non-Hermitian models has been inspired by Bessis \cite{DB} and by
Bender and Boettcher \cite{BB} who revealed that one of important
non-Hermitian models in field theory {\em seemed} to possess the
discrete and real ``bound-state-like" spectrum. This hypothesis
(which has rigorously been confirmed a few years later \cite{DDT})
opened the question of the possible physical interpretation of the
model because its $H_{\it e\!ff}$ only happened to satisfy a
weaker form of eq. (\ref{quasihermitian}),
 \be
 H_{\it e\!ff}^\dagger={\cal P}\,
 H_{\it e\!ff}\,{\cal P}^{-1}\,,\ \ \ \ \ \ \ \
 {\cal P}^{-1}={\cal P}={\cal P}^\dagger,
 \label{pseudohermitian}
 \ee
with the role of an indefinite ``pseudo-metric" played by the
operator of parity~\cite{conservation}.

In the latter new family of models (conventionally called, for
some historical reasons \cite{BG}, ${\cal PT}-$symmetric) people
only succeeded in constructing physical metric $\Theta$ very
recently \cite{Jones}. One must remember that $H_{\it e\!ff}\neq
H_{\it e\!ff}^\dagger$ so that the standard Schr\"{o}dinger
equation must be considered together with the parallel
Hermitian-conjugate problem,
 \be
 H_{\it e\!ff} |\psi\rangle = E\,|\psi\rangle,
 \label{SE}
 \ \ \ \ \ \ \ \ \ \
 H_{\it e\!ff}^\dagger |\psi\rangle\rangle =
 E\,|\psi\rangle\rangle.
 \label{SEc}
 \ee
Moreover, many ${\cal PT}-$symmetric models proved to behave, in
many a respect, against our current intuition \cite{anomalous}.
Hence, up to a few partial differential exceptions \cite{Calog},
people prefer working with the one-dimensional effective
Hamiltonians. Hence, it is not too surprising that the
constructions of the correct positive definite metric $\Theta$
found their most transparent presentations in exactly solvable
models. For example, one may recollect ref. \cite{Batal} where an
entirely elementary ``schematic" square-well model of
refs.~\cite{sqw} has been used and studied and where its
measurable and physical aspects have been described in detail.

Our present letter was immediately motivated by some specific
features of the transition ${\cal P}\to \Theta$ in the context of
coupled-channel Schr\"{o}dinger equations, perceived here as
residing somewhere in between the unsolvable (=``realistic") and
solvable extremes. Their introduction may be based on various
physical assumptions as well as on some formal considerations in
representation theory \cite{JakubskydiplomaOndra} and/or on
constructions belonging to relativistic quantum mechanics
\cite{polub} etc.

\section{Coupled-channel problems}

In some of the contemporary papers devoted to ${\cal
PT}-$symmetric Quantum Mechanics \cite{BBjmp} the operator ${\cal
P}$ of eq. (\ref{pseudohermitian}) does not coincide with parity
and it need not even be chosen as involutive. For this reason, let
us now change its symbol from ${\cal P}$, say, to $\theta$.
Although the rigorous mathematical specification of this
$\theta\neq \theta^{-1}$ need not be easy in general
\cite{Kretschmer}, a generic requirement is that this auxiliary
indefinite pseudo-metric operator as well as the related
Hamiltonian $H_{\it e\!ff}$ remain sufficiently elementary
\cite{Ventura}.

\subsection{Two particles coupled in a one-dimensional deep box}

The most elementary coupled-channel model are, undoubtedly, the
two-channel models and, in particular, their one-dimensional
example
 \be
 H_{\it e\!ff} =\left (
 \begin{array}{cc}
 -\frac{d^2}{dx^2}&0\\
0&
 -\frac{d^2}{dx^2}
 \ea
 \right ) + V_{\it e\!ff},
 \ \ \ \ \ \ \
 V_{\it e\!ff} =\left (
 \begin{array}{cc}
 V_a(x)&W_b(x)\\
 W_a(x)&V_b(x)
 \ea
 \right )
 \label{vazanekanaly}
 \ee
written in units $\hbar = 2m = 1$. These operators lie somewhere
in between the ordinary and partial differential ones and offer a
certain combination of merits of the solvability (so
characteristic for ordinary differential equations) with a richer
structure of wave functions.

In what follows we shall only search for such two-channel bound
states  $|\psi\rangle$ which have the standard asymptotic
Dirichlet boundary conditions mimicked by their suitable large$-L$
approximation
 \be
 \langle \pm L|\psi\rangle =\left (
 \ba
 \langle \pm L|\psi_a\rangle\\
 \langle \pm L|\psi_b\rangle
 \ea
 \right )=\left (
 \ba
 0\\0
 \ea
 \right ).
 \label{bc}
 \ee
In this setting, a non-trivial core of our message will lie in the
assumption that while the character of the spectrum $\{ E_n\}$
will be assumed ``entirely standard" (i.e., real, discrete and
bounded below), the potential term itself will possess an unusual,
asymmetric and manifestly {\em non-Hermitian} form with $V_a \neq
V_a^\dagger$, etc.

\subsection{Models with $\theta-$pseudo-Hermiticity
{\em and} ${\cal PT}-$symmetry
}

Our key motivation stems form the success of a number of the
ordinary differential non-Hermitian Hamiltonians which were
offered within the framework of the so called quasi-Hermitian
\cite{Geyer}, ${\cal PT}-$symmetric \cite{BB} or ${\cal
P}-$pseudo-Hermitian \cite{ali} Quantum Mechanics. In this
context, rather surprisingly, coupled-channel models were not yet
studied in sufficient detail. Our present letter has been written
just to fill this gap.

For the sake of a maximal transparency of our forthcoming
arguments we shall violate the current Hermiticity as drastically
as possible and postulate the $\theta-$pseudo-Hermiticity property
 \be
 H_{\it e\!ff}^\dagger=\theta\,
 H_{\it e\!ff}\,\theta^{-1}\,\ \ \ \ \ \ \theta=\theta^\dagger
 \label{thetapseudohermitian}
 \ee
using the following parity-dependent (or, if you wish,
generalized-parity-dependent) pseudo-metric
 \be
 \theta =\theta^\dagger=\left (
 \begin{array}{cc}
 0&{\cal P}\\
 {\cal P}&0
 \ea
 \right ), \ \ \ \ \
 \theta^{-1}=\left (
 \begin{array}{cc}
 0&{\cal P}^{-1}\\
 {\cal P}^{-1}&0
 \ea
 \right )
 .
 \label{provazanekanaly}
 \ee
As long as these operators commute with the kinetic (i.e.,
differential) part of our $H_{\it e\!ff}$ (\ref{vazanekanaly}),
the related $\theta-$pseudo-Hermiticity condition
(\ref{thetapseudohermitian}) degenerates to the following two
${\cal P}-$pseudo-Hermiticity relations and one definition,
 \be
 W^\dagger_a={\cal P}W_a{\cal P}^{-1}
 , \ \ \ \ \ \ \ \
 W^\dagger_b={\cal P}W_b{\cal P}^{-1}
 , \ \ \ \ \ \ \ \
 V_b={\cal P}^{-1}V_a^\dagger {\cal P}
 \,.
 \label{potencosh}
 \ee
In our present note just a ``minimal" model will be considered,
with the re-scaled coordinate $x$ [such that $L=1$ in
eq.~(\ref{bc})] and with the current parity operator such that
${\cal P} \varphi(x) = \varphi (-x)$.

\section{Exactly solvable square-well example \label{3.2}}

In one of the simplest versions of the above Hamiltonian
(\ref{vazanekanaly}) let us consider the following purely
imaginary square-well realization of the off-diagonal ${\cal
P}-$pseudo-Hermitian potentials,
  \be
 \ba
 {\rm Re}\,W_{a,b}(x) =0, \ \ \ \ \ x \in (-1,1),\\
 {\rm Im}\,W_a(x) =Z, \ \ \ \ \ {\rm Im}\,W_b(x) =Y, \ \ \ \ \
  x \in (-1,0),\\
 {\rm Im}\,W_a(x) =-Z, \ \ \ \ \ {\rm Im}\,W_b(x) =-Y, \ \ \ \ \ x \in
 (0,1),
 \ea
 \label{SQW}
 \ee
accompanied by the trivial intra-channel interactions,
$V_a=V_b=0$.

\subsection{Trigonometric wave functions}

The obvious ansatz
 \be
 \ba
 \varphi(x)=\langle x|\psi_a\rangle= \left \{
 \begin{array}{ll}
  A\,\sin \kappa_L(x+1)
 , \ \ & x \in (-1,0), \\
 C\,\sin \kappa_R(1-x),
  \ \ & x \in (0,1), \ \ \
 \ea
 \right .
 \\
 \chi(x)=\langle x|\psi_b\rangle= \left \{
 \begin{array}{ll}
  B\,\sin \kappa_L(x+1)
 , \ \ & x \in (-1,0), \\
 D\,\sin \kappa_R(1-x),
  \ \ & x \in (0,1), \ \ \
 \ea
 \right .
 \ea
 \label{ansatzf}
 \ee
may be normalized as usual, with $\varphi(x)=\varphi^*(-x)$ and
$\chi(x)=\chi^*(-x)$ giving $C=A^*$, $D=B^*$ and
$\kappa_L=\kappa_R^*=\kappa=s-{\rm i}\,t$ where, say, $s>0$. Its
insertion in the differential Schr\"{o}dinger eqs.~(\ref{SE})
leads to the complex solvability condition
 \be
 \left(
 \begin{array}{cc}
 \kappa^2-E& {\rm i}Z\\
 {\rm i}Y& \kappa^2-E
 \ea
 \right )
 \left (
 \ba
 A\\B
 \ea
 \right )=0.
 \label{vosum}
 \ee
The related complex secular equation may be re-read as two real
conditions,
 \be
 \left \{
 \begin{array}{llll}
 2st=\pm \sqrt{YZ},\ & E = s^2-t^2 & \ {\rm for}\ & YZ > 0,\\
% & {\rm sign}\,t =\pm 1,
  t=0, & E = s^2 \pm  \sqrt{-YZ} &\ {\rm for}\ & YZ < 0.
 \ea
 \right .
 \ee
We note that our matrix problem (\ref{vosum}) is in fact Hermitian
at $Y = -Z$. Hence, let us only study the more challenging former
option with $Y>0$, $Z>0$ in what follows.

\subsection{Matching conditions at $x=0$}

What we have to postulate is the continuity of both the wave
functions $\varphi(x)$ and $\chi(x)$ and of their first
derivatives at $x=0$. Two of the resulting four complex equations
 \ben
 \ba
 A \sin \kappa =
 A^* \sin \kappa^*,\ \ \ \ \ \ \ \ \
 A \kappa \cos \kappa =
 -A^* \kappa^* \cos \kappa^*,\\
 B \sin \kappa =
 B^* \sin \kappa^*,\ \ \ \ \ \ \ \ \
 B \kappa \cos \kappa =
 -B^* \kappa^* \cos \kappa^*,
 \ea
 \een
specify the ratio of coefficients $A/B = 2st/Y$ as real. The
remaining two equations
 \be
 \left (
 \begin{array}{cc}
 \sin \kappa& -\sin \kappa^*\\
 \kappa\cos \kappa& \kappa^*\cos \kappa^*
 \ea
 \right )
 \left (
 \ba
 A\\
 A^*
 \ea
 \right )=0
 \ee
define one of these complex coefficients. The nontriviality of
this solution is guaranteed by the elementary secular equation
${\rm Re}\,(\kappa^{-1}\tan \kappa)=0$. The later condition has
the simplified equivalent form
 \be
 s\,\sin 2s + t\,\sinh 2t = 0
 \label{secular}
 \ee
with the structure of solutions known from the single-well
constructions~\cite{sqwb},
 \be
 s=s_n=\frac{(n+1)\pi}{2}+(-1)^n\varepsilon_n
 , \ \ \ \ \ n = 0, 1, \ldots
 \ee
where quantities $\varepsilon_n$ remain small and positive at
large $n$ or small $\sqrt{YZ}$. In contrast to the single-well
case, the present real energy levels $E_n=s^2_n-YZ/(4s_n^2)$ are
doubly degenerate since $t$ in the above-mentioned ratio $A/B=
2st/Y = \pm \sqrt{Z/Y}$ may acquire both signs,
 \be
 |\psi_n^{(\sigma)}\rangle =\left (
 \begin{array}{l}
 |\varphi_n\rangle \cdot \sqrt{Z} \\
 |\varphi_n\rangle \cdot \sigma\,\sqrt{Y}
 \ea
 \right ), \ \ \ \ \  \ \ \sigma = \pm 1, \ \ \ \  n = 0, 1, \ldots\,.
 \label{bcmod}
 \ee
The construction is completed.

\section{Discussion}

\subsection{The existence of the set of two commuting observables \label{3}}

The key merit of our choice of the example (\ref{SQW}) is
methodical since its Hamiltonian $H_{\it e\!ff}$ commutes with the
operator which might play the role of an independent spin-like
observable in our system,
 \be
 \Omega=
 \left (
 \begin{array}{cc}
 0& \sqrt{Z/Y}\\
 \sqrt{Y/Z}&0
 \ea
 \right ).
 \label{spin}
 \ee
Indeed, the $\theta-$pseudo-Hermiticity $\Omega^\dagger = \theta
\Omega\theta^{-1}$ is readily verified as one of the welcome
intuitive arguments supporting the possible consistency of such an
interpretation.

We saw that {\em both} our pseudo-Hermitian candidates $H=H_{\it
e\!ff}$ and $\Omega$ for observables possess the real spectra
(remember: just two points $\sigma = \pm 1$ in the latter case),
at not too large couplings $Y\geq 0$ and $Z \geq 0$ at least (more
precisely, at all of them such that $\sqrt{YZ} < Z_{crit} \approx
4.48$ \cite{sqw}). In this regime, our wave functions may be
perceived as functions of the (real) variables $E$ (or,
equivalently, $n$) and $\sigma$. From such a point of view their
set becomes complete once the operators in question (i.e., $H$ and
$\Omega$ in our model) form a {\em complete set of commuting
operators} in a given indefinite metric.

\subsection{The basis in Hilbert space for the degenerate spectrum}

In the majority of studies concerned with ${\cal PT}-$symmetric
quantum mechanics the energy spectra happen to be non-degenerate.
In contrast, bound states are usually classified by more quantum
numbers in practice~\cite{Geyer,time}. In this sense, the
degeneracy of energies and the emergence of the second quantum
number $\sigma$ might further enhance the pragmatic as well as
theoretical appeal of our present example.

In the present unusual non-Hermitian setting, the eigenvectors of
$H^\dagger$ and $\Omega^\dagger$ (or, equivalently, the left
eigenvectors of $H$ and $\Omega$) will be also needed. We must
solve the following extended set of Schr\"{o}dinger equations,
 \ben
 H\,|E,\sigma\rangle = E\,|E,\sigma\rangle, \ \ \ \ \ \
 \Omega\,|E,\sigma\rangle = \sigma\,|E,\sigma\rangle,
 \een
 \ben
 \langle\langle E,\sigma|\,H=E\,\langle\langle E,\sigma|
 , \ \ \ \ \ \
 \langle\langle E,\sigma|\,\Omega=\sigma\,\langle\langle
  E,\sigma|
 \,. \
 \een
Fortunately, the latter pair only means that
 \ben
 H^\dagger\,|E,\sigma\rangle\rangle = E^*\,|E,\sigma\rangle\rangle
 , \ \ \ \ \ \
 \Omega^\dagger\,|E,\sigma\rangle\rangle =
  \sigma^*\,|E,\sigma\rangle\rangle
 \een
so that, due to the pseudo-Hermiticity
(\ref{thetapseudohermitian}) and due to the independence and
completeness of our set of wave functions (be it proved or
assumed) we have
 \be
 |E,\sigma\rangle\rangle =\theta\,
 |E^*,\sigma^*\rangle\,q_{E\sigma}\,, \ \ \ \ \ E=E_1, E_2, \ldots,
 \ \ \ \ \sigma = \pm 1.
 \label{star}
 \ee
We are just left with a freedom in a complex normalization
constant in the explicit definition of all the missing solutions.

Now, it is easy to derive the biorthogonality relations among our
wave functions,
 \ben
 \langle\langle E',\sigma'|E,\sigma\rangle(E'-E)=0, \ \ \ \ \ \
 \langle\langle E',\sigma'|E,\sigma\rangle(\sigma'-\sigma)=0.
 \een
We see that only the diagonal overlaps may remain non-vanishing
and enter the completeness relations
 \ben
 I = \sum_{E\!,\,\sigma}\
 |E,\sigma\rangle\,\frac{1}
 {\langle\langle E,\sigma|E,\sigma\rangle}
 \langle\langle E,\sigma|
 \een
as well as the following two spectral representation formulae,
 \be
 H = \sum_{E\!,\,\sigma}\
 |E,\sigma\rangle\,\frac{E}
 {\langle\langle E,\sigma|E,\sigma\rangle}
 \langle\langle E,\sigma|, \ \ \ \ \ \ \
 \Omega = \sum_{E\!,\,\sigma}\
 |E,\sigma\rangle\,\frac{\sigma}
 {\langle\langle E,\sigma|E,\sigma\rangle}
 \langle\langle E,\sigma|
 \label{spectral}
 \ee
where the later one is rather formal of course.

\section{The transition from indefinite $\theta$ to physical
$\Theta$}

In terms of the metric $\Theta$, the bound-state coupled-channel
wave functions of our model acquire the standard probabilistic
interpretation. Indeed, the scalar product
 \be
  \left (
 |\psi_1\rangle \odot |\psi_2\rangle
 \right )
 =
 \langle \psi_1|\, \Theta\,|\psi_2 \rangle
 =
 \langle \psi_1 |\psi_2\rangle_{(physical)} \,
 \ee
generates the norm, $||\psi||=\sqrt{\langle \psi
|\psi\rangle_{(physical)}}$, and enables us to treat all the
{quasi-Hermitian} operators $A$ with the property
$A^\dagger=\Theta\,A\,\Theta^{-1}$ as observables. Such a usage of
this word makes good sense because the expectation values $\langle
\psi|\,A\, |\psi\rangle_{(physical)}$ are mathematically
unambiguously defined,
 \be
\left (
 |\psi_1\rangle \odot |A\,\psi_2\rangle
 \right )
 \equiv
  \left (
 |A\,\psi_1\rangle \odot |\psi_2\rangle
 \right )\,.
 \ee
In our particular square-well model just a re-interpretation of
our $\theta-$pseudo-Hermitian Hamiltonian $H_{\it e\!ff}$ and spin
$\Omega$ as quasi-Hermitian operators with respect to $\Theta$ is
needed.

\subsection{A formula for the metric}

Let us recollect that we started our considerations from a given
(i.e., with a strong preference, {\em very} simple) indefinite
metric (i.e., pseudo-metric) operator $\theta$ and from a
$\theta-$pseudo-Hermitian Hamiltonian $H_{\it e\!ff}$ [cf. eq.
(\ref{thetapseudohermitian})]. Now, having performed all the
constructions of the bound states we are left with the ultimate
task of finding the {\em physical metric}, i.e., a Hermitian and
positive definite {\em solution} $\Theta = \Theta^\dagger > 0$ of
eq.~(\ref{quasihermitian}). In terms of the above formulae
(\ref{spectral}) it is easy to see, immediately, that we must have
 \ben
 \Theta = \sum_{E\!,\,\sigma\!,\,F\!,\,\tau}\
 |F,\tau\rangle\rangle\ R_{F\!,\tau\!,\,E\!,\,\sigma}\
 \langle\langle E,\sigma| \
 \een
where the (in general, fairly ambiguous \cite{Geyer}) choice of
the matrix $R$ must remain compatible with
eq.~(\ref{quasihermitian}) and with its analogue for $\Omega$.
This remains true if and only if
 \ben
  R_{E\!,\sigma\!,\,F\!,\,\tau}\
 \left (E^*-F\right )=0, \ \ \ \ \
  R_{E\!,\sigma\!,\,F\!,\,\tau}\
 \left (\sigma-\tau \right )=0.
 \een
Once the spectrum of energies is assumed real we arrive at the
compact formula
 \be
 \Theta = \sum_{E\!,\,\sigma}\
 |E,\sigma\rangle\rangle\ S_{E\!,\,\sigma}\
 \langle\langle E,\sigma| \
 \label{capitaltheta}
 \ee
which represents the menu of {\em all} the eligible pseudo- and
metrics parametrized by the infinite sequence of the non-vanishing
parameters $S_{E\!,\,\sigma}$ where $\sigma = \pm 1$ and, by
assumption, $E=E_0, E_1, \ldots$ are all real. Easily we also
deduce that
 \ben
 \Theta^{-1} = \sum_{E\!,\,\sigma}\
 |E,\sigma\rangle\
 \frac{1/S_{E\!,\,\sigma}}{
 \langle E,\sigma|E,\sigma\rangle\rangle\cdot
  \langle\langle E,\sigma|E,\sigma\rangle
 }\
 \langle E,\sigma| \ .
 \een
The obligatory invertibility and Hermiticity of $\Theta$ is
guaranteed when all the parameters $S_{E\!,\,\sigma}$ remain real
and non-vanishing. Finally, its positivity (i.e., tractability as
a physical metric) will be achieved whenever all
$S_{E\!,\,\sigma}$ remain positive.

\subsection{Quasi-parity}

In our particular model of section \ref{3.2}, the reality of the
energies was comparatively easy to prove. In such a situation,
people usually work with eq. (\ref{star}) and employ very
particular $q=q_{E\sigma}=\pm 1$, calling such a ``new quantum
number" quasi-parity \cite{conservation,ptho} or charge
\cite{BBJ}. It enters the formula
 \ben
 \langle\langle E,\sigma|E,\sigma\rangle=q_{E\sigma}
 \langle E,\sigma|\,\theta\,
 |E,\sigma\rangle\,, \ \ \ \ \ E=E_1, E_2, \ldots,
 \ \ \ \ \sigma = \pm 1,
 \ \ \ \ \ \
 q_{E\sigma}=\pm 1.
 \een
In our model where the proportionality of both the components of
our wave functions $|\psi_n^{(\sigma)}\rangle$ to the same
single-channel ket $|\varphi_n\rangle$ is a useful artifact, we
may insert eqs. (\ref{provazanekanaly}) and (\ref{bcmod}) and
arrive at an even more compact relation
 \ben
 \langle\langle \psi_n^{(\sigma)}|\psi_n^{(\sigma)}
 \rangle=q_{E_n \sigma}\cdot
 \langle \psi_n^{(\sigma)}|\,\theta\,
 |\psi_n^{(\sigma)}\rangle
 =
 \sigma\ q_{E_n \sigma}\cdot
 \langle \varphi_n|\,{\cal P}\,
 | \varphi_n\rangle \cdot \sqrt{4YZ}
 \,.
 \een
Obviously, we may {\em prescribe} the overall sign of this overlap
since it is controlled

\begin{itemize}

\item by $\sigma= \pm 1$, i.e., by the optional sign-convention
accepted in eq. (\ref{bcmod}),

\item by the overlap $\langle \varphi_n|\,{\cal P}\,|
\varphi_n\rangle$ which ``measures" the parity of the
upper-channel wave function in eq. (\ref{ansatzf})  and {\em
varies} with $n = 0, 1, \ldots$,

\item and by the quasi-parity $q_{E_n\sigma}=\pm 1$ which is our
free choice.

\end{itemize}

 \noindent
Due to the mere two-by-two matrix character of the spin $\Omega$,
our key definition (\ref{capitaltheta}) of the metric may be now
further reduced to the sum
 \be
 \Theta = \sum_{n=0}^\infty\
 {\cal P}\,
 |\varphi_n\rangle
 \left (
 \begin{array}{cc}
 Y\,\left (S_{E_n\!,\,+}+S_{E_n\!,\,-}\right )
 &
 \sqrt{YZ}\,\left (S_{E_n\!,\,+}-S_{E_n\!,\,-}\right )\\
 \sqrt{YZ}\,\left (S_{E_n\!,\,+}-S_{E_n\!,\,-}\right )
 &
 Z\,\left (S_{E_n\!,\,+}+S_{E_n\!,\,-}\right )
 \ea
 \right )
 \langle\varphi_n| \,
 {\cal P}\,
 \label{capitalthetanove}
 \ee
which is a two-by-two matrix with respect to the spin (\ref{spin})
and depends on the pairs of the positive free parameters
$S_{E_n\!,\,\pm}>0$.

\section{Summary and outlook}

\subsection{An efficiency of perturbation expansions}

One of the key advantages of our present model is that the
coordinate representation (\ref{ansatzf}) of its wave functions is
piecewise trigonometric. In addition, perturbation ansatz of the
form
 \ben
 \varepsilon_n = \sum_{k=1}^K
 \left [\frac{YZ}{(n+1)^2\pi^2}
 \right ]^k\  \cdot \ \sum_{t=1}^{T(k)} \frac{c_{k,t}}{(n+1)^t\pi^t}
 \een
may be used to solve eq. (\ref{secular}) by iterations. Then, for
the sufficiently high excitations $n \geq n_0 \gg 1$ and/or for
the sufficiently small geometric-mean measure $\sqrt{YZ}$ of the
non-Hermiticity of our $H_{\it e\!ff}$ we may derive and work,
say, with the formula
 \be
 \varepsilon_n =\left [
 \frac{2\,YZ}{(n+1)^3\pi^3}
  +
 \frac{4\,Y^2Z^2}{3\,(n+1)^5\pi^5}\right ]\,\left [1
 +{\cal O}\left (\frac{1}{(n+1)^4}\right )\right ]
 +{\cal O}\left (\frac{Y^3Z^3}{(n+1)^7}\right )
 \label{firsttwo}
 \ee
showing that the convergence in $1/(n+1)$ proves extremely rapid.

It would be easy to demonstrate that one of the important
consequences of the steady growth of the latter quantities with
growing $YZ$ would be a merger and the subsequent complexification
of $s_0$ and $s_1$ (or of $s_2$ and $s_3$ etc) at a sufficiently
large $YZ$. At this critical point (or rather critical curve
$Y=const/Z$), the reality of the spectrum of our
parity-pseudohermitian $H_{\it e\!ff}$ gets spontaneously broken.
This possibility  opens a number of questions which were not
discussed here at all.

Another immediate consequence of eq. (\ref{firsttwo}) is that the
role of the non-Hermiticity decreases very quickly at the higher
excitations. This means that in our present model a fairly
reliable approximation  of the metric  $\Theta_{(approx)}$ will
sufficiently significantly differ from the unit operator
$\Theta_{(trivial)}=I$ just in a {\em finite-dimensional} subspace
spanned, say, by the $N$ lowest excitations of the $Y=Z = 0$
system of the two completely decoupled deep (and, of course,
Hermitian) square wells.

\subsection{Point-interaction models}

In the future, our present choice and study of our example could
prove insufficient. Then, it need not necessarily be followed just
by its various generalizations with a suitable piece-wise form of
the forces. Indeed, with the number of the admissible
discontinuities, one may expect an increase of difficulties of a
purely technical nature. We believe that at least a partial
reduction of such a obstacle could be achieved when one switches
to the class of the point interactions, say, of the ${\cal
PT}-$symmetric form
 \be
 W_a(x) =\sum_{\ell=1}^{M_a}
 \left [
 i\,\alpha_{\ell}\,\delta \left (x-a_{\ell}\right )-
 i\,\alpha_{\ell}\,\delta \left (x+a_{\ell}\right )
 \right ]
 \,,
 \label{potika}
 \ee
 \be
 W_b(x) =\sum_{j=1}^{M_b}
 \left [
 i\,\beta_{j}\,\delta \left (x-b_{j}\right )-
 i\,\beta_{j}\,\delta \left (x+b_{jl}\right )
 \right ]
 \,,
 \label{potikb}
 \ee
 \be
 V_a(x) = V_b^*(-x) =\sum_{n=1}^{N}
 i\,\gamma_{n}\,\delta \left (x-g_{n}\right )
 \,
 \label{potikc}
 \ee
with the purely imaginary delta functions  located at certain
ordered sets of the points
 \ben
0 < a_1 < \ldots <a_{M_a}<1, \ \ \ 0 < b_1 < \ldots <b_{M_b}<1, \
\ \  -1 < g_1 < \ldots <g_{N}<1
 \een
and proportional to some real constants $\alpha_\ell$,  $\beta_j$
and $\gamma_n$.

\section*{Acknowledgement}

Work partially supported by the grant Nr. A 1048302 of GA AS CR.

%\end{document}

\end{document}